\documentclass[a4paper,10pt,nofootinbib]{revtex4}

\UseRawInputEncoding
\usepackage{caption}
\usepackage{graphicx} % standard LaTeX graphics tool;
\usepackage{amsmath} % For getting proper math eqns
\usepackage{amssymb} % Used for getting various symbols eg. gtrsim, lesssim etc.
\usepackage{bm} % For bold math (esp of lower greek letters)
\usepackage{dcolumn}% Align table columns on decimal point
\usepackage{color}
\usepackage{mathrsfs}
\usepackage{amsfonts}
\usepackage{varioref}
\usepackage{mathrsfs}
\usepackage{graphicx}
\usepackage{latexsym}
\usepackage{amsmath}
\usepackage{amssymb}
\usepackage{textcomp}
\usepackage{amsbsy}
\usepackage{graphics}
\usepackage{epstopdf}
\usepackage{color}
\usepackage[caption=false]{subfig}
 
\RequirePackage[colorlinks,citecolor=blue,urlcolor=magenta,linkcolor=blue]{hyperref}
\input epsf

\allowdisplaybreaks[4]
\begin{document}
\tolerance=5000

\title{Generalized entropic dark energy with spatial curvature}

\author{Sergei~D.~Odintsov$^{1,2}$\,\thanks{odintsov@ieec.uab.es},
Simone~D'Onofrio$^{2}$\,\thanks{donofrio@ice.csic.es},
Tanmoy~Paul$^{3}$\,\thanks{pul.tnmy9@gmail.com}} \affiliation{
$^{1)}$ ICREA, Passeig Luis Companys, 23, 08010 Barcelona, Spain\\
$^{2)}$ Institute of Space Sciences (ICE, CSIC) C. Can Magrans s/n, 08193 Barcelona, Spain\\
$^{3)}$ Department of Physics, Visva-Bharati University, Santiniketan -731235, India.}

%\date{}

\tolerance=5000

\begin{abstract}
In the realm of thermodynamics of apparent horizon, we construct a dark energy (DE) model from 4-parameter generalized entropy of apparent horizon in a spatially non-flat universe. In particular, considering a non-zero spatial curvature of the universe, we determine the dark energy fractional density and the dark energy equation of state (EoS) parameter (corresponding to the 4-parameter generalized entropy) in closed analytic forms. It turns out that the scenario can describe the correct thermal history of the universe, with the sequence of matter and dark energy epochs. Comparing with the $\Lambda$CDM model, the proposed generalized entropic DE model provides a higher value of present Hubble parameter for certain range of entropic parameter(s) leading to a possible resolution of Hubble tension issue. This in turn leads to a positive spatial curvature of the universe. We confront the scenario with CC \& BAO, Pantheon+ \& SH0ES and joint analysis of the CC \& BAO \&  Pantheon+ \& SH0ES datasets, which clearly depicts the phenomenological viability of the present model for some best fitted values of entropic parameter(s) that are indeed consistent with the resolution of Hubble tension.
\end{abstract}

\maketitle

\section{Introduction}
In the context of thermodynamics of apparent horizon, the cosmological field equations are derived from the first law of thermodynamics applied on the apparent horizon, with a certain form of entropy of the horizon \cite{Jacobson_1995,Cai:2005ra,Cai:2006rs,Nojiri:2023wzz} (for a recent review on entropic cosmology, see \cite{Nojiri:2024zdu}). In this regard, the Bekenstein-Hawking like entropy of the horizon leads to the usual Friedmann equations. Consequently it has been recently showed that Einstein gravity (cosmology) naturally validates the second law of thermodynamics of the apparent horizon \cite{Odintsov:2024ipb}, which makes the inter connection between gravity and thermodynamics more concrete. However a different form of horizon entropy than the Bekenstein-Hawking one results to modified Friedmann equations which may have rich cosmological consequences. Some well known non-additive entropies that are extensively used in cosmology are Tsallis entropy \cite{Tsallis:1987eu}, the R\'{e}nyi entropy \cite{Renyi}, the Barrow entropy \cite{Barrow_2020}, the Sharma-Mittal entropy \cite{Sayahian_Jahromi_2018}, the Kaniadakis entropy \cite{Kaniadakis_2005}, the Loop Quantum gravity entropy \cite{Majhi:2017zao} etc. All of these entropies prove to be function of the Bekenstein-Hawking entropy variable ($S$) and share the following properties: (a) monotonically increases with respect to the Bekenstein-Hawking entropy and (b) vanishes in the limit $S \rightarrow 0$. Owing to these common properties, one may naturally ask the question like --- ``does there exist a generalized entropy that can bring all the known entropies proposed so far within
a single umbrella?'' Motivated by this, a 4-parameter generalized entropy has been proposed in \cite{Nojiri_2022}, which is able to generalize all the aforementioned entropies for suitable representations of generalized entropic parameters (for some other forms of generalized entropy with larger number of parameters, see \cite{Nojiri:2022aof,Odintsov:2022qnn}). Here it may be mentioned that according to the conjecture stated in \cite{Nojiri_2022}, the minimum number of parameters required for a generalized entropy is equal to four. In cosmological context, it consequently becomes of considerable interest to constrain the four parameters present in the generalized entropy from various perspectives. It turns out that the generalized entropy has rich consequences towards the early universe cosmology (particularly from inflation to reheating as well as in the context of bouncing scenario) as well as to the black hole physics, and the corresponding constraints on the entropic parameters are addressed from different perspectives \cite{Nojiri:2022aof,Nojiri_2022,Odintsov:2022qnn,Odintsov_2023,Bolotin:2023wiw,Lymperis:2023prf,doi:10.1142/S0219887824503389,Odintsov:2024sbo}. On other hand, from the usual thermodynamic perspective, the generalized entropies have a microscopic interpretation too --- in both the canonical and grand-canonical descriptions, the generalized entropies can be interpreted as the statistical ensemble average of a series of microscopic quantity(ies) given by various powers of $\left(-k\ln{\rho_\mathrm{e}}\right)^{n}$ (with $n$ being a positive integer and $\rho_\mathrm{e}$ symbolizes the phase space density of the respective ensemble), along with a term representing the fluctuation of Hamiltonian and number of particles of the system under consideration (in case of canonical ensemble, the fluctuation on the particle number vanishes) \cite{Nojiri:2023bom}. Importantly, as we will show that for the constraints on the generalized entropic parameters, required to have a viable dark energy era and concomitantly resolve the Hubble tension (see Table.~[\ref{table:Best_Fit}]), the generalized entropy does not tend to any of the known entropies like the Tsallis entropy \cite{Asghari:2021lzu}, the Barrow entropy \cite{Leon:2021wyx,Asghari:2021bqa}, the Kaniadakis entropy \cite{Hernandez-Almada:2021rjs} etc. The above arguments clearly depict the importance of generalized entropy in the field of cosmology as well as in black hole physics.\\
Cosmological observations \cite{riess1998observational, perlmutter1999measurements,arnaud2016planck, ahn2012ninth} have provided good evidence for the late time acceleration of the universe, but its true nature still remains unknown. It is generally expected that the dark energy component should constitute about 70\% of the total energy budget of the universe, in order to have sufficient negative pressure and can produce the desired acceleration of the universe. Several dark energy models have been constructed, like, the cosmological constant model ($\Lambda$CDM), holographic dark energy model \cite{Bousso_2002, fischler1998, Nojiri_2006, Saridakis:2020zol, Sinha_2020, Adhikary:2021}, dynamical dark energy models \cite{RatraPRD1988, CaldwellPRL2003, Bamba_2012,Nojiri_2005, Yang_2017, Yang_2019} etc., but no single theory can consistently describe the dark energy era. For instance, the $\Lambda$CDM model perfectly describes the dark energy during the late universe, except for the fact that the $\Lambda$CDM model is plagued with the Hubble tension. Such a tension arises due to the difference in the measured value of the present Hubble parameter $H_\mathrm{0}$ from different sources. In particular, the cosmic microwave background (CMB) data from the Planck satellite along with Baryon Acoustic Oscillation (BAO) data \cite{Planck2020, BAO2017, BAO2011}, Big Bang Nucleosynthesis (BBN) \cite{BBN2021}  and Dark Energy Survey (DES) \cite{DES12018, DES22018, DES:2017tss}  have constrained the value to be $H_\mathrm{0} \sim (67.0 - 68.5)$km/s/Mpc, whereas the measurement from the SH0ES, TRGB and H0LiCOW collaborations \citep{Sh0ES2019, H0LiCOW2019} predict the value to be $H_\mathrm{0} \sim (74.03 \pm 1.42)$ km/s/Mpc. Such anomaly on $H_\mathrm{0}$ from different sources can not be concomitantly addressed by the $\Lambda$CDM model \citep{Abdalla:2022yfr} and is an open problem in cosmology. This makes the true nature of dark energy as one of the important question in modern cosmology and thus the search for dark energy candidate is still on. In this regard several attempts have been made to resolve the tension from different directions \cite{Vagnozzi:2023nrq,Odintsov:2020qzd,Khalife:2023qbu,Verde:2019ivm,Knox:2019rjx,DiValentino:2020zio,DiValentino:2021izs,Perivolaropoulos:2021jda,Pedrotti:2024kpn,Yang:2018uae,Mamon_2017,Niedermann:2020dwg,DiValentino:2019jae,DiValentino:2020naf,ROY2022101037}.
In the current work, we are interested to construct a dark energy candidate through the thermodynamic route of apparent horizon in a spatially non-flat universe, particularly from the 4-parameter generalized entropy of the horizon. The motivations behind such set-up are the following: (a) the 4-parameter generalized entropy is the minimal version of generalized entropy, (b) beside the entropic parameters, the model can also constrain the spatial curvature of the present universe (if any), and (c) the 4-parameter generalized entropic model deviates from the $\Lambda$CDM one and thus the deviation possibly addresses the Hubble tension.\\
The paper is organized as follows: after describing the basic cosmological field equations corresponding to the 4-parameter generalized entropy in a spatially non-flat universe in Sec.~[\ref{sec-I}], we determine various relevant quantities of dark energy era in Sec.~[\ref{sec-Ia}]. The next Sec.~[\ref{sec-II}] is reserved for the data analysis and the results. The paper ends with some concluding remarks in Sec.~[\ref{Sec-conclusion}].

\section{Generalized entropy and the corresponding Friedmann equations in spatially non-flat scenario}\label{sec-I}

In this section, we will derive the cosmological field equations from 4-parameter generalized entropy in spatially non-flat case. In this regard, the spatially non-flat FLRW spacetime has the metric as,
\begin{equation}
    ds^2 = -dt^2 + a^2(t)\left(\frac{dr^2}{1-k r^2} + r^2 d\Omega^2\right) \ ,
\end{equation}
where $t$ is the cosmic time, $a(t)$ is the scale factor of universe, $k$ represents the curvature parameter and $d\Omega^2$ symbolizes the line element of a 2-sphere having unit radius. For the above metric, the radius of the apparent horizon takes the following form \cite{Cai_2005}
\begin{equation}\label{Horizon-Non-Flat}
    r_\mathrm{A} = \frac{1}{\sqrt{H^2 + k/a^2}} \ ,
\end{equation}
with $H = \dot{a}/a$ being the Hubble parameter of universe. Having the $r_\mathrm{A}$ in hand, we now derive the cosmological field equations from the thermodynamics of the apparent horizon given by:
\begin{equation}
 TdS_\mathrm{h} = -d\left(\rho V\right) + WdV \ ,
 \label{n1}
\end{equation}
where $T = \frac{1}{2\pi r_\mathrm{A}} \left|1 - \frac{\dot{r_\mathrm{A}}}{2Hr_\mathrm{A}}\right|$ and $S_\mathrm{h}$ represent the temperature and the entropy, respectively, associated to the horizon. Moreover, the quantities in the rhs of Eq.~(\ref{n1}) are for the matter fields inside the horizon, in particular, $W = \frac{1}{2}\left(\rho - p\right)$ is known as the work density term where $\rho$ and $p$ are the energy density and the pressure of the matter fields inside the horizon. For a general form of horizon entropy: $S_\mathrm{h}=S_\mathrm{h}(S)$ (where $S = A/(4G)$ is the Bekenstein-Hawking entropy and $A = 4\pi r_\mathrm{A}^2$ is the area of the horizon), the first law of thermodynamics of the apparent horizon leads to \cite{odintsov2023entropicinflationpresencescalar,Nojiri:2024zdu},
\begin{equation}\label{Gen-Thermo-Eq}
   \frac{\Dot{r}_\mathrm{A}}{r_\mathrm{A}^3} \frac{\partial S_\mathrm{h}}{\partial S}  = - \frac{4 \pi G }{3}\Dot{\rho} \ .
\end{equation}
Owing to Eq.~(\ref{Horizon-Non-Flat}), one can get
\begin{equation}
    \Dot{r}_\mathrm{A} = - H r_\mathrm{A}^3 \left(\Dot{H} - \frac{k}{a^2}\right) \ ,
\end{equation}
which, along with Eq.~(\ref{Gen-Thermo-Eq}), yields
\begin{equation}\label{Gen-Thermo-Eq-with-Horizon}
     H  \left(\Dot{H}-\frac{k}{a^2}\right)\frac{\partial S_\mathrm{h}}{\partial S}  = \frac{4 \pi G }{3}\Dot{\rho} \ .
\end{equation}
With the local energy conservation of the matter fields: $\Dot{\rho}+3H(\rho+p)=0$, we find the second Friedmann equation in spatially non-flat case:
\begin{equation}\label{Gen_Fried_II}
     \left(\Dot{H}-\frac{k}{a^2}\right)\frac{\partial S_\mathrm{h}}{\partial S}  = -4\pi G(\rho+p) \ .
\end{equation}
The first Friedmann equation can be obtained by integrating both sides of Eq.~(\ref{Gen-Thermo-Eq-with-Horizon}), and as a result, we get
\begin{equation}\label{Gen_Fried_I}
    \int d\left(\frac{1}{r_\mathrm{A}^2}\right)\frac{\partial S_\mathrm{h}}{\partial S}  = \frac{8 \pi G }{3}\rho + \frac{\Lambda}{3} \ ,
\end{equation}
where the cosmological constant $\Lambda$ naturally arises as an integrating constant. Thus, as a whole, Eq.~(\ref{Gen_Fried_I}) and Eq.~(\ref{Gen_Fried_II}) are the Friedmann equations derived from the thermodynamics of the apparent horizon in the spatially non-flat scenario where $S_\mathrm{h}$ represents the entropy of the horizon. In the present work, we consider the horizon entropy to be the 4-parameter generalized entropy given by \cite{Nojiri_2022},
\begin{equation}\label{gen-entropy}
    S_\mathrm{h} \equiv S_\mathrm{g}(\alpha_\mathrm{\pm},\beta,\gamma)= \frac{1}{\gamma}\left[\left(1+\frac{\alpha_+}{\beta}S\right)^\beta-\left(1+\frac{\alpha_-}{\beta}S\right)^{-\beta}\right] \ ,
\end{equation}
with recall that $S = A/(4G)$ is the Bekenstein-Hawking like entropy; and $\alpha_\mathrm{\pm}$, $\beta$, and $\gamma$ are the entropic parameters which are assumed to be positive in order to have a monotonic increasing function of $S_\mathrm{g} = S_\mathrm{g}(S)$. For the 4-parameter generalized entropy, the Friedmann Eq.~(\ref{Gen_Fried_I}) and Eq.~(\ref{Gen_Fried_II}) take the following forms,
\begin{equation}\label{Fried-Eq-1}
    \begin{aligned}
    \frac{G\beta x^2 }{\pi \gamma} & \left[ \frac{1}{2+\beta}\left(\frac{G\beta x}{\pi\alpha_-}\right)^{\beta} 2F_1\left(1+\beta,2+\beta,3+\beta,-\frac{G\beta x}{\pi\alpha_-}\right)\right. \\ 
    & \left.+ \frac{1}{2-\beta}\left(\frac{G\beta x}{\pi\alpha_+}\right)^{-\beta} 2F_1\left(1-\beta,2-\beta,3-\beta,-\frac{G\beta x}{\pi\alpha_+}\right) \right] = \frac{8 \pi G}{3}\rho + \frac{\Lambda}{3} \ , 
\end{aligned}
\end{equation}
and
\begin{equation}\label{Fried-Eq-2}
     \left(\Dot{H}-\frac{k}{a^2}\right) \frac{1}{\gamma}\left[\alpha_+\left(1+\frac{\pi\alpha_+}{\beta G x}\right)^{\beta-1}+\alpha_-\left(1+\frac{\pi\alpha_-}{\beta G x}\right)^{-\beta-1}\right]=-4\pi G(\rho+p) \ ,
\end{equation}
respectively, where for convenience we introduced the variable $x \equiv 1/r_\mathrm{A}^2 = H^2 + k/a^2$. Introducing the curvature fractional density parameter $\Omega_k \equiv k/a^2H^2$, we have $x = H^2 ( 1+\Omega_k)$. The above two equations can be equivalently expressed by,
\begin{align}\label{DE-FRW}
    H^2+\frac{k}{a^2} &= \frac{8\pi G}{3}\left(\rho+\rho_\mathrm{g}\right) + \frac{\Lambda}{3} \\
    \Dot{H} -\frac{k}{a^2} &= -4\pi G \left[\left(\rho+\rho_\mathrm{g}\right)+\left(p+p_\mathrm{g}\right)\right] \ ,
\end{align}
where $\rho_\mathrm{g}$ and $p_\mathrm{g}$ have the following forms:
\begin{equation}\label{entropy-energy}
    \begin{aligned}
        \rho_\mathrm{g} = \frac{3}{8\pi G} \Bigg\{ x - \frac{G\beta x^2 }{\pi \gamma} & \left[ \frac{1}{2+\beta}\left(\frac{G\beta x}{\pi\alpha_-}\right)^{\beta} 2F_1\left(1+\beta,2+\beta,3+\beta,-\frac{G\beta x}{\pi\alpha_-}\right)\right. \\
     & \left.+ \frac{1}{2-\beta}\left(\frac{G\beta x}{\pi\alpha_+}\right)^{-\beta} 2F_1\left(1-\beta,2-\beta,3-\beta,-\frac{G\beta x}{\pi\alpha_+}\right) \right] \Bigg\}
    \end{aligned}
\end{equation}
and
\begin{equation}\label{entropy-pressure}
     p_\mathrm{g} = \frac{\Dot{H} - k/a^2}{4\pi G}\left\{\frac{1}{\gamma}\left[\alpha_+\left(1+\frac{\pi\alpha_+}{\beta G x}\right)^{\beta-1}+\alpha_-\left(1+\frac{\pi\alpha_-}{\beta G x}\right)^{-\beta-1}\right] -1\right\} - \rho_\mathrm{g}\ ,
\end{equation}
respectively. Here $\rho_\mathrm{g}$ and $p_\mathrm{g}$ arises from the 4-parameter generalized entropy and thus they may be called as entropic energy density and entropic pressure. The presence of such $\rho_\mathrm{g}$ and $p_\mathrm{g}$ certainly modify the cosmological field equations compared to the usual ones, which may have rich cosmological consequences during different epochs of universe.

At this stage it deserves mentioning that the four parameter generalized entropy has a gravitational perspective, in particular, the generalized entropy admits a Lagrangian description from a modified theory of gravity. Recently some of our authors proposed a direct correspondence between modified gravity and entropic cosmology \cite{Nojiri:2025fiu,Nojiri:2025gkq}. This reveals how to determine the equivalent entropy of the apparent horizon (in the sector of entropic cosmology) for a given modified gravity, or, vice-versa. It turns out that the four parameter generalized entropy can be equivalently represented by $F(Q)$ (or $F(T)$) gravitational Lagrangian given by (here $Q = -6H^2$ is the non-metricity scalar and $T = -6H^2$ is the torsion scalar; \cite{Heisenberg:2023lru}):
\begin{eqnarray}
 F(Q) = 3\left(-Q\right)^{1/2}\int^{-\frac{Q}{6}}~dy~(6y)^{-\frac{3}{2}}~\int^{y}dx\left(\frac{\partial S_\mathrm{g}}{\partial S}\right)\bigg|_{S = \frac{\pi}{Gx}} \, ,
 \label{new-1}
\end{eqnarray}
(see \cite{Nojiri:2025fiu}). Using the expression of $S_\mathrm{g}$ from Eq.~(\ref{gen-entropy}) to the above expression, one gets the following form of $F(Q)$:
\begin{eqnarray}
 F(Q) = \frac{3G\beta}{\gamma}\left(-Q\right)^{1/2}\int^{-\frac{Q}{6}}~dy~(6)^{-\frac{3}{2}}~y^{\frac{1}{2}}\bigg[\frac{1}{(2+\beta)}\left(\frac{G\beta y}{\pi \alpha_{-}}\right)^{\beta}2F_1\left(1+\beta,2+\beta,3+\beta,-\frac{G\beta y}{\pi\alpha_-}\right)\nonumber\\
 + \frac{1}{(2-\beta)}\left(\frac{G\beta y}{\pi\alpha_+}\right)^{-\beta} 2F_1\left(1-\beta,2-\beta,3-\beta,-\frac{G\beta y}{\pi\alpha_+}\right)\bigg] \, .
 \label{new-2}
\end{eqnarray}
Eq.~(\ref{new-2}) shows the form of $F(Q)$ corresponding to the four parameter generalized entropy. The above expression also gives the desired form of $F(Q)$ for the other horizon entropies, like the Tsallis entropy, the R\'{e}nyi entropy, the Sharma-Mittal entropy, the Kaniadakis entropy etc., by taking suitable limits of the parameters $\alpha_{\pm}$, $\beta$ and $\gamma$ \cite{Nojiri:2025fiu}.

As mentioned in the introduction, we are mainly interested on the late time dark energy era of the universe in the present work, in which case, the Hubble parameter is much less than the Planck scale and thus we can safely consider $GH^2 \ll 1$. Owing to this condition, Eq.~(\ref{Gen_Fried_I}) and Eq.~(\ref{Gen_Fried_II}) becomes,
\begin{equation}
    H^2 (1+\Omega_k) = \left[(2-\beta)\frac{\gamma}{\alpha_+}\left(\frac{\beta G }{\pi \alpha_+}\right)^{\beta-1}\left(\frac{8 \pi G}{3}\rho + \frac{\Lambda}{3}\right)  \right] ^ \frac{1}{2-\beta} 
\end{equation}
and 
\begin{equation}
    \frac{\alpha_+}{\gamma}\left(\frac{\beta G H^2}{\alpha_+\pi}(1+\Omega_k)\right)^{1-\beta}\left(\Dot{H}-\frac{k}{a^2}\right) = - 4\pi G (\rho + p) \ ,
\end{equation}
respectively, where the leading order terms of $GH^2$ are retained. It may be noted that the standard Friedmann equations are recovered for $\beta = 1$ and $\gamma = \alpha_{+}$, this is however expected as the 4-parameter generalized entropy converges to the Bekenstein-Hawking like entropy for such parameter choices, which leads to the standard Friedmann equations. Moreover, in the limit $G H^2 \ll 1$, the $\rho_\mathrm{g}$ and $p_\mathrm{g}$ (from Eq.~(\ref{entropy-energy}) and Eq.~(\ref{entropy-pressure})) become,
\begin{equation}\label{Equiv_Fluid_Quantities_Approx}
    \rho_\mathrm{g} = \frac{3x}{8\pi G} \left(1-\frac{\sigma}{2-\beta} x^{1-\beta}\right) \hspace{1.cm}\text{and}\hspace{1.cm} p_\mathrm{g} =- \frac{\Dot{H}-k/a^2}{4 \pi G} \left(1-\sigma \, x^{1-\beta}\right) - \rho_g \ ,
\end{equation}
by introducing $\sigma \equiv \frac{\alpha_+}{\gamma}\left(\frac{G\beta}{\pi\alpha_+}\right)^{1-\beta}$ as the combination of entropic parameters.

\subsection{Dark Energy era}\label{sec-Ia}
In this section, we will concentrate on late time cosmological implications of the 4-parameter generalized entropy in spatially non-flat case. Eq.~(\ref{DE-FRW}) acts as the main governing equation where the dark energy (DE) density is contributed from the entropic energy density ($\rho_\mathrm{g}$) and the cosmological constant ($\Lambda$); and moreover, $\rho$ is sourced from the pressureless dust and the radiation as well, i.e. $\rho = \rho_\mathrm{m} + \rho_\mathrm{R}$. However due to $\rho_\mathrm{R} \ll \rho_\mathrm{m}$ at late time, we consider $\rho \approx \rho_\mathrm{m}$. The dark energy density and the dark pressure are given by,
\begin{equation}\label{DE-energy density}
    \rho_\mathrm{D} \equiv \rho_\mathrm{g} + \rho_\Lambda = \rho_\mathrm{g} + \frac{3}{8\pi G}\left(\frac{\Lambda}{3}\right) \ ,
\end{equation}
and
\begin{equation}\label{DE-pressure}
    \rho_\mathrm{D} + p_\mathrm{D} = \rho_\mathrm{g} + p_\mathrm{g} \ ,
\end{equation}
where $\rho_\mathrm{g}$ and $p_\mathrm{g}$ are obtained in Eq.~(\ref{entropy-energy}) and Eq.~(\ref{entropy-pressure}) respectively.
Using such expressions of $\rho_\mathrm{g}$ and $p_\mathrm{g}$, the equation of state parameter for the DE is found to be,
\begin{equation}\label{w_D_General}
    w_\mathrm{D} = \frac{p_\mathrm{D}}{\rho_\mathrm{D}} = -1 - \frac{2\left(\Dot{H}-k/a^2\right)\left(1-
\sigma x^{1-\beta}\right)}{\Lambda+3x\left(1-\frac{\sigma}{2-\beta}x^{1-\beta}\right)} \ .
\end{equation}
The above expression of $w_\mathrm{D}$ indicates that the spatially flat $\Lambda$CDM scenario can be recovered with  $\beta = 1$, $\gamma = \alpha_+$ and $k = 0$, for which $w_\mathrm{D} = -1$. This is expected because for such choices of the entropic parameters, $\rho_\mathrm{g}$ and $p_\mathrm{g}$ vanish and thus the present cosmic scenario becomes identical with the spatially flat $\Lambda$CDM model. Actually, a different form of horizon entropy than the Bekenstein-Hawking one, in particular, the 4-parameter generalized entropy leads to a non-zero $\rho_\mathrm{g}$ and $p_\mathrm{g}$ which differs the model from the $\Lambda$CDM one. This has some interesting implications, as we will show that the 4-parameter generalized entropy can lead to a viable dark energy era in spatially non-flat scenario and concomitantly resolve the Hubble tension issue.

With the above ingredients in hand, the Friedmann equation is given by,
\begin{equation}
    H^2 + \frac{k}{a^2} = \frac{8\pi G}{3} \left(\rho_\mathrm{m} + \rho_\mathrm{D}\right) \ .
\end{equation}
Let us introduce the fractional density parameters for individual energy components as: $\Omega_k \equiv k/a^2H^2$, $\Omega_\mathrm{m} \equiv\frac{8 \pi G}{3 H^2}\rho_\mathrm{m}$ and $\Omega_\mathrm{D} \equiv\frac{8 \pi G}{3 H^2}\rho_\mathrm{D}$. Then the above equation is equivalently written as,
\begin{equation}\label{I_Fried_Fract}
    \Omega_\mathrm{m} + \Omega_\mathrm{D} - \Omega_\mathrm{k} = 1 \ ,
\end{equation}
which has to be accompanied with the conservation equations:
\begin{align}
    \Dot{\rho}_\mathrm{m} + 3H\rho_\mathrm{m} &= 0 \\
    \Dot{\rho}_\mathrm{D} + 3 H \rho_\mathrm{D}\left(1 + w_\mathrm{D}\right) &=0 \ .
\end{align}
Using these conservation equations we can link the fractional density parameters to their values at present time, denoted with the suffix “0", as
\begin{equation}
    \Omega_\mathrm{m} = \frac{8 \pi G}{3 H^2}\rho_\mathrm{m} = \frac{8 \pi G}{3 H^2}\rho_\mathrm{m0}\left(\frac{a_\mathrm{0}}{a}\right)^3 = \Omega_\mathrm{m0}\frac{H_\mathrm{0}^2 a_\mathrm{0}^3}{H^2 a^3}
\end{equation}
and 
\begin{equation}
    \Omega_\mathrm{k} = \frac{k}{a^2H^2} = \frac{k}{a_\mathrm{0}^2H_\mathrm{0}^2} \frac{a_\mathrm{0}^2H_\mathrm{0}^2}{a^2 H^2} = \Omega_\mathrm{k0}\left(\frac{a_\mathrm{0}H_\mathrm{0}}{a H}\right)^2 \ ,
\end{equation}
where we defined $\Omega_\mathrm{m0} \equiv \frac{8 \pi G}{3 H_\mathrm{0}^2}\rho_\mathrm{m0}$ and $\Omega_\mathrm{k0} \equiv \frac{k}{a_\mathrm{0}^2 H_\mathrm{0}^2}$. Plugging the above expressions into Eq.~(\ref{I_Fried_Fract}) immediately leads to,
\begin{equation}\label{I_Fried_(z)}
     1- \Omega_D(z) = \frac{1}{H^2}\left(H_0^2(z+1)^3\Omega_{m0}-H_0^2(z+1)^2\Omega_{k0}\right) \ ,
\end{equation}
in terms of red shift factor defined by $z = \frac{a}{a_\mathrm{0}} - 1$. With a little bit of rearrangement, Eq.~(\ref{I_Fried_(z)}) provides the Hubble parameter (in terms of $z$) as follows,
\begin{equation}\label{H}
    H(z) = \frac{H_\mathrm{0}(z+1)}{\sqrt{1-\Omega_\mathrm{D}(z)}}\sqrt{(z+1)\Omega_\mathrm{m0}-\Omega_\mathrm{k0}} \ .
\end{equation}
Differentiating both sides with respect to cosmic time ($t$), and using $\Dot{z} = - H (1+z)$, we find
\begin{equation}
    \Dot{H} = -\frac{H^2}{2(1-\Omega_\mathrm{D}(z))} \left[3(1-\Omega_\mathrm{D}(z)) + (1+z)\frac{d\Omega_\mathrm{D}}{dz} + \frac{H_\mathrm{0}^2}{H^2}(1+z)^2\Omega_\mathrm{k0} \right] \ ,
\end{equation}
Consequently, the deceleration parameter ($q$) is obtained as
\begin{equation}\label{dec-parameter}
    q \equiv -1 -\frac{\Dot{H}}{H^2} = -1 + \frac{1}{2(1-\Omega_\mathrm{D})}\left[3(1-\Omega_\mathrm{D})+(1+z)\frac{d\Omega_\mathrm{D}}{dz}+\frac{\left(1-\Omega_\mathrm{D}\right)\Omega_\mathrm{k0}}{(z+1)\Omega_\mathrm{m0}-\Omega_\mathrm{k0}}\right] \ .
\end{equation}
Another useful quantity to characterize the evolution of the universe is the DE EoS ($w_\mathrm{D}$, see Eq.~(\ref{w_D_General})), for which,
we need the following expression for $x$ (recall that the quantity $x$ is introduced after Eq.~(\ref{Fried-Eq-2})),
\begin{equation}\label{Def_x}
    x \equiv H^2(1+ \Omega_k) =\frac{H_0^2(z+1)^2}{1-\Omega_D}\left[(z+1)\Omega_{m0}-\Omega_{k0}\Omega_D\right] \ ,
\end{equation}
Owing to the above expression of $x$, the DE EoS parameter from Eq.~(\ref{w_D_General}) is found to be,
\begin{equation}\label{eos-parameter}
    w_\mathrm{D} = -1 +\frac{\mathscr{N}}{\mathscr{D}}
\end{equation}
with
\begin{equation}
    \begin{aligned}
        \mathscr{N} \equiv \frac{H_\mathrm{0}^2(z+1)^2}{(1-\Omega_\mathrm{D})^2}\left[(z+1)\Omega_\mathrm{m0}-\Omega_\mathrm{k0}\right] &\left\{3(1+\Omega_\mathrm{D})+(1+z)\frac{d\Omega_\mathrm{D}}{dz}+\frac{(3-2\Omega_\mathrm{D})(1-\Omega_\mathrm{D})}{(z+1)\Omega_\mathrm{m0}-\Omega_\mathrm{k0}}\Omega_\mathrm{k0}\right\} \\
        &\times \left\{1-\sigma\left[\frac{H_\mathrm{0}^2(z+1)^2}{1-\Omega_\mathrm{D}}[(z+1)\Omega_\mathrm{m0}-\Omega_\mathrm{k0}\Omega_\mathrm{D}]\right]^{1-\beta} \right\}
    \end{aligned}
\end{equation}
and 
\begin{equation}
        \mathscr{D} \equiv \Lambda + \frac{3H_\mathrm{0}^2(z+1)^2}{1-\Omega_\mathrm{D}}[(z+1)\Omega_\mathrm{m0}+\Omega_\mathrm{k0}\Omega_\mathrm{D}]\left\{1-\frac{\sigma}{2-\beta}\left[\frac{H_\mathrm{0}^2(z+1)^2}{1-\Omega_\mathrm{D}}[(z+1)\Omega_\mathrm{m0}-\Omega_\mathrm{k0}\Omega_\mathrm{D}]\right]^{1-\beta} \right\} \ ,
\end{equation}
respectively. It is evident that both the deceleration parameter and the dark energy EoS parameter depend on $\Omega_\mathrm{D}(z)$ and its derivative. Therefore in order to have $w_\mathrm{D} = w_\mathrm{D}(z)$ and $q = q(z)$, i.e. in terms of redshift factor, we need to determine the dark energy density parameter in terms of $z$. The explicit calculations for $\Omega_\mathrm{D}= \Omega_\mathrm{D}(z)$ and its derivative are reported in Appendix \ref{app:Omega_D}, in particular,
\begin{equation}\label{A8}
    \Omega_\mathrm{D}(z) = 1+ \left\{\frac{\Omega_\mathrm{k0}}{\Tilde{\Omega}(z)}-\left[\frac{1+\frac{\Omega_\mathrm{k0}}{\Tilde{\Omega}(z)}+\frac{\Lambda}{3H_0^2(1+z)^2\Tilde{\Omega}(z)}}{\frac{\sigma}{2-\beta}\left[H_\mathrm{0}^2(1+z)^2\Tilde{\Omega}(z)\right]^{1-\beta}}\right]^{\frac{1}{2-\beta}}\right\}^{-1} \ ,
\end{equation}
where $\Tilde{\Omega}(z) = (z+1)\Omega_\mathrm{m0}-\Omega_\mathrm{k0}$. The above expression at $z = 0$ provides a constraint relation between $\left\{\Lambda,H_\mathrm{0},\Omega_\mathrm{m0},\Omega_\mathrm{k0}\right\}$, and is given by,
\begin{equation}\label{constraint}
    \Lambda = \frac{3\sigma}{2-\beta}\left[H_\mathrm{0}^2(1+\Omega_\mathrm{k0})\right]^{2-\beta}- 3 H_0^2\Omega_\mathrm{m0} \ .
\end{equation}
This relation can be then used to eliminate the cosmological constant from the equations and study the system with a parameter less. Thus, as a whole, Eq.~(\ref{dec-parameter}), Eq.~(\ref{eos-parameter}) and Eq.~(\ref{A8}) provide the analytic expressions of deceleration parameter, DE EoS parameter and the DE fractional density parameter in the context of 4-parameter generalized entropy for a spatially non-flat scenario. It may be realized that $q(z)$, $w_\mathrm{D}(z)$ and $\Omega_\mathrm{D}(z)$ depend on the entropic parameters $\beta$ and $\sigma$ along with observationally determined parameters $H_\mathrm{0}$, $\Omega_\mathrm{m0}$ and $\Omega_\mathrm{k0}$ (recall the constraint relation (\ref{constraint}) that relates $\Lambda$ with $\left\{H_\mathrm{0},\Omega_\mathrm{m0},\Omega_\mathrm{k0}\right\}$). In particular, in the next section we will see how some of the latest data on the expansion history of our universe can fix the set of parameters $(H_0, \Omega_{m0}, \Omega_{k0}, \sigma)$, and consequently, can help in alleviating the tension on the estimation of $H_\mathrm{0}$.
It may be noted that, out of the four parameters present in the generalized entropy, only two parameters ($\beta$ and $\sigma$) are relevant to represent the dark energy quantities; actually, the other entropic parameters get encapsulated within $\sigma$ (see after Eq.~(\ref{Equiv_Fluid_Quantities_Approx})). This is important from the fact that the maximum number of parameters for a physical entropy is two \cite{Hanel_2011}, which is indeed consistent with the present context where, effectively, two entropic parameters ($\beta$ and $\sigma$) are left to get all the dark energy quantities. Moreover, from the perspective of information theory, the four parameter generalized entropy should obey the Kolmogorov-Nagumo axioms, which aim to extend the classical concept of entropy to accommodate different forms of uncertainty and disorder in complex systems \cite{Jizba_2020}. The validity of such axioms has been proved for some of the entropies present in the literature, such as the Tsallis and the R\'{e}nyi ones \cite{Corominas_Murtra_2024}. In this way, the four parameter generalized entropy respects such axioms in some limit of the parameters. It will be the subject of future works, more devoted to the statistical side of the subject, to show that the validity of such axioms is respected by the generalized entropy.

\section{Data analysis with observational datasets and results}\label{sec-II}

Data analysis is a significant feature in cosmology, which provides the best fitted values of the model parameters from different datasets. In this section, we present a brief description of the various observational datasets used and the methodology adopted to constrain the parameters of the model. In particular, here we will consider the datasets of Cosmic chronometers (CC), Baryon Acoustic Oscillations (BAO) and Pantheon+ respectively, along with their joint analysis. We perform the standard Bayesian analysis \cite{padilla2021cosmological} to obtain the posterior distribution of the parameters by employing a Markov Chain Monte Carlo (MCMC) method, and for this purpose, we use the publicly available \texttt{emcee} library package in Python \cite{foreman2019emcee} to carry out the MCMC analysis.

\subsection{Cosmic chronometers (CC)}
Comparing the age of galaxies it is possible to obtain  the expansion rate $H(z)$ at a given redshift  \cite{Jimenez_2002}. From the definition of the redshift, the Hubble parameter is linked to the differential age $dz/dt$ by
\begin{equation}
    H(z) = - \frac{1}{1+z}\frac{dz}{dt} \ . 
\end{equation}
Taking two ensembles of galaxies formed at the same time at a relatively small redshift difference ($dz$), it is possible to get the difference in cosmic time ($dt$) by comparing the ages of the two galaxies. The estimation on the age of the galaxy is done using  stellar population models and spectroscopy \cite{Simon_2005,Ratsimbazafy_2017}. The galaxy evolution sets cosmic chronometers at different redshifts, which can then be used to obtain the expansion history. \\
For the CC data, we used 31 data points lying in the redshift range $ 0.07 \leq z \leq 1.965 $, the data are given in the  left side of Table \ref{table:CC_Data}. The corresponding $\chi^2$ function is given by
\begin{equation}\label{chi_CC}
    \chi^2_{\text{CC}} = \sum_{i=1}^{31}\frac{(H^{obs}(z_i)-H^{th}(z_i))^2}{\sigma^2_H(z_i)} \ ,
\end{equation}
where $H^{obs}(z_i)$ is the observed value with uncertainty $\sigma_H(z_i)$, and $H^{th}(z_i)$ represents the theoretical prediction given by Eq.~(\ref{H}).
We consider the observations at various redshift to be uncorrelated and no systematic covariance matrix has been introduced as it does not produce considerable change in the CC data analysis. However in the case of the joint analysis of different datasets, we incorporate the covariance matrix  \cite{Moresco_2020}.

\subsection{Baryon Acoustic Oscillations (BAO)}
Primordial perturbations produced in the early universe acoustic sound waves propagating through the plasma. This plasma was formed by photons and baryons, which then lost this highly coupled state after the decoupling era. This waves where traveling trough the plasma creating denser and sparser zones. Studying the two-point galaxy correlation function it can be seen a peak at a certain preferred distance, corresponding to the distance traveled by the sound wave until the decoupling $r_d$, at which there is an over density of galaxies due to the sound wave. This pattern can be observed from the CMB pattern or from galaxy clusters at different redshifts. From the second method we can obtain an information of the expansion rate studying how much the standard ruler $r_d$ has been stretched due to the universe expansion. \\
For the BAO data we used 26 data points lying in the redshift range $ 0.24 \leq z \leq 2.36 $, the data are given in right side of Table \ref{table:CC_Data}. Also in this case we have no cross correlations terms and the $\chi^2$ function is equivalent to Eq.~(\ref{chi_CC}).

\setlength{\arrayrulewidth}{0.1mm}
\setlength{\tabcolsep}{4pt}
\renewcommand{\arraystretch}{1.2}

\begin{table}[htb]
\centering
\begin{tabular}{|c | c | c | c | c | c | c | c |} 
 \hline
 \multicolumn{8}{|c|}{CC} \\
 \hline \hline
 $z$ & $H(z)$ & $\sigma_H$ & Refs & $z$ & $H(z)$ & $\sigma_H$ & Refs \\ 
\hline
$0.070$ & $69$ & $19.6$ & \cite{zhang2014four} & $0.4783$ & $80.9$ & $9$ &  \cite{Moresco_2016} \\
$0.090$ & $69$ & $12$ & \cite{Simon_2005} & $0.480$ & $97$ & $62$ & \cite{Daniel_Stern_2010}\\
$0.120$ & $68.6$ & $26.2$ & \cite{zhang2014four} & $0.593$ & $104$ & $13$ & \cite{M_Moresco_2012}\\
$0.170$ & $83$ & $8$ &  \cite{Simon_2005} & $0.6797$ & $92$ & $8$ & \cite{M_Moresco_2012}\\
$0.1791$ & $75$ & $4$ & \cite{M_Moresco_2012} & $0.7812$ & $105$ & $12$ & \cite{M_Moresco_2012}\\
$0.1993$ & $75$ & $5$ & \cite{M_Moresco_2012} & $0.8754$ & $125$ & $17$ & \cite{M_Moresco_2012}\\
$0.200$ & $72.9$ & $29.6$ & \cite{zhang2014four} & $0.880$ & $90$ & $40$ & \cite{Daniel_Stern_2010}\\
$0.270$ & $77$ & $14$ & \cite{Simon_2005}  & $0.900$ & $117$ & $23$ &  \cite{Simon_2005} \\
$0.280$ & $88.8$ & $36.6$ & \cite{zhang2014four} & $1.037$ & $154$ & $20$ & \cite{M_Moresco_2012}\\
$0.3519$ & $83$ & $14$ & \cite{M_Moresco_2012} & $1.300$ & $168$ & $17$ &  \cite{Simon_2005}\\
$0.3802$ & $83$ & $13.5$ & \cite{Moresco_2016} & $1.363$ & $160$ & $33.6$ &\cite{Moresco_2015} \\
$0.400$ & $95$ & $17$ &  \cite{Simon_2005}  & $1.430$ & $177$ & $18$ &   \cite{Simon_2005}\\
$0.4004$ & $77$ & $10.2$ &  \cite{Moresco_2016} & $1.530$ & $140$ & $14$ & \cite{M_Moresco_2012}\\
$0.4247$ & $87.1$ & $11.2$ &  \cite{Moresco_2016} & $1.750$ & $202$ & $40$ & \cite{M_Moresco_2012}\\
$0.4497$ & $92.8$ & $12.9$ &  \cite{M_Moresco_2012} & $1.965$ & $186.5$ & $50.4$ & \cite{Moresco_2015}\\
$0.470$ & $89$ & $34$ &  \cite{Ratsimbazafy_2017} &   &   &   & \\
\hline
\end{tabular} 
\hspace{.7cm}
\begin{tabular}{|c | c | c | c | c | c | c | c |} 
 \hline
 \multicolumn{8}{|c|}{BAO} \\
 \hline \hline
 $z$ & $H(z)$ & $\sigma_H$ & Refs & $z$ & $H(z)$ & $\sigma_H$ & Refs \\ 
\hline
$0.24$ & $79.69$ & $2.99$ &  \cite{Gazta_aga_2009} & $0.57$ & $96.8$ & $3.4$ & \cite{Anderson_2014}\\
$0.3$ & $81.7$ & $6.22$ & \cite{Oka_2014} & $0.59$ & $98.48$ & $3.18$ &\cite{Wang_2017}\\
$0.31$ & $78.18$ & $4.74$ & \cite{Wang_2017} & $0.60$ & $87.9$ & $6.1$ & \cite{Blake_2012}\\
$0.34$ & $83.8$ & $3.66$ & \cite{Gazta_aga_2009} & $0.61$ & $97.3$ & $2.1$ & \cite{Alam_2017}\\
$0.35$ & $82.7$ & $9.1$ & \cite{Chuang_2013} & $0.64$ & $98.82$ & $2.98$ & \cite{Wang_2017}\\
$0.36$ & $79.94$ & $3.38$ & \cite{Wang_2017} & $0.73$ & $97.3$ & $7.0$ & \cite{Blake_2012}\\
$0.38$ & $81.5$ & $1.9$ &  \cite{Alam_2017}  & $2.30$ & $224$ & $8.6$ & \cite{Busca_2013}\\
$0.40$ & $82.04$ & $2.03$ & \cite{Wang_2017} & $2.33$ & $224$ & $8$ & \cite{Bautista_2017}\\
$0.43$ & $86.45$ & $3.97$ & \cite{Gazta_aga_2009} & $2.34$ & $222$ & $8.5$ & \cite{Delubac_2015}\\
$0.44$ & $82.6$ & $7.8$ & \cite{Blake_2012} & $2.36$ & $226$ & $9.3$ & \cite{Font_Ribera_2014}\\
$0.44$ & $84.81$ & $1.83$ & \cite{Wang_2017} &  &  &  & \\
$0.48$ & $87.79$ & $2.03$ & \cite{Wang_2017} &  &  &  & \\
$0.51$ & $90.4$ & $1.9$ & \cite{Alam_2017} &  &  &  & \\
$0.52$ & $94.35$ & $2.64$ & \cite{Wang_2017} &   &  & & \\
$0.56$ & $93.34$ & $2.3$ &\cite{Wang_2017} & &  &  & \\
$0.57$ & $87.6$ & $7.8$ &\cite{Chuang_2016}  &   &   &   & \\

\hline
\end{tabular}
\caption{Values of $H(z)$ at different redshift and its uncertainty obtained for the two different sources, the CC and the BAO.}
\label{table:CC_Data}
\end{table}

\subsection{Supernova Type Ia (SNIa)}
The first prove of the late time acceleration of the universe was due to the study of supernovae \cite{Riess_1998,Perlmutter_1999} which are powerful explosion of stars and thus can represent good point sources for estimating cosmic distances due to their huge luminosity. Among all the types, the Supernovae of Type Ia represent the most luminous and homogeneous kind. SNIa are supernovae that occurs in a binary system in which one of the companions is a white dwarf. They are called standard candles since a relation between the brightness peak and the observer distance can be archived  which, along with the estimation of the redshift, can characterize the expansion history of the universe. The most up-to-date catalog of SNIa is the \textit{Pantheon+} dataset \cite{Brout_2022}, which consist of 1624 data points. These data can be extended using the SH0ES \cite{Riess_2022} results, which relies on the imaging of Cepheid variable stars in the host galaxies of recent, nearby SNIa. This leads to a joint dataset of 1701 data points spanning in the redshift range $0.01\leq z\leq 2.3$ \footnote{Data can be found at the following link: \url{https://github.com/PantheonPlusSH0ES/DataRelease} .}.
Each data point is given by the observed distance modulus at a given redshift $\mu^{obs}(z)$. The $\chi^2$ function in this case is
\begin{equation}
    \chi^2_{\text{SNIa}} = \sum_{i,j=1}^{1701}(\mu^{obs}(z)-\mu^{th}(z))_i\Big(\text{Cov}^{-1}\Big)_{ij}(\mu^{obs}(z)-\mu^{th}(z))_j \ ,
\end{equation}
where $\text{Cov}$ is the covariant matrix of the dataset obtained summing the statistical and systematic covariance matrices. The theoretical expression of the distance modulus is 
\begin{equation}
\mu^{th}(z) = 5 \, \log_{10}\left(\frac{d_L(z)}{10 pc}\right)+25 \ ,
\end{equation}
where the link with the DE model is given by the luminosity distance $d_L(z)$ defined as
\begin{equation}
    d_\mathrm{L}(z) = (1+z) \int_0^z\frac{dz'}{H(z')} \ .
\end{equation}
Clearly $d_\mathrm{L}(z)$ encodes the information about the cosmological dynamics through the presence of the Hubble parameter, and in the present analysis, we will the expression of $H(z)$ from Eq.~(\ref{H}).

\subsection*{Results}
In order to find the best fitted values of the model parameters we use the standard approach of Bayesian analysis with a Monte Carlo Markov Chain (MCMC) method, implemented using the \texttt{emcee} package in Python \cite{Foreman_Mackey_2019}. We study the model given by the Hubble parameter in Eq. (\ref{H}) along with the dark energy fractional density parameter obtained in \eqref{Omega_D(z)}, corresponding to the 4-parameter generalized entropy in spatially non-flat universe. Clearly $H(z)$ as well as $\Omega_\mathrm{D}(z)$ depend on the entropic parameters $\left\{\beta,\sigma\right\}$ and the observationally determined quantities $\left\{H_0,\Omega_{m0},\Omega_{k0}\right\}$. As mentioned earlier that the current entropic dark energy model converges to the $\Lambda$CDM case for the parametric choices given by: $\beta = \sigma =1$. However the model needs to be constructed in such a way that it has a small variation than the $\Lambda$CDM one at late times in order to to have a possible resolution of the Hubble tension issue. In particular, our aim is to check the viability of the proposed generalized entropic DE model ensuring that its deviation from the $\Lambda$CDM model is minimal. There are two ways by which one can constrain the parameter space such that the deviation between the proposed model and the $\Lambda$CDM model is minimal: either $\beta = 1$ and vary $\sigma$ or vice versa. However it has been found that the case with $\sigma = 1$ and varying $\beta$ cannot depict a correct cosmological scenario, and thus we consider $\beta = 1$ with varying $\left\{H_0,\Omega_{m0},\Omega_{k0},\sigma\right\}$ which indeed predicts a consistent cosmological evolution of the universe, as we will depict below. We perform the MCMC with flat priors for these parameters in the ranges $H_0 \in [50,90]$, $\Omega_{m0}\in [0.27,0.33]$, $\Omega_{k0}\in [-0.1,0.25]$ and $\sigma \in [0.6,1.3]$ \footnote{For the CC \& BAO dataset the prior on  $\Omega_{k0}$ has ben chosen to be $\Omega_{k0}\in [-0.1,0.1]$, since the dataset does not seem to provide a constrain for this parameter, this prior has been changed in the SNIa analysis and in the joint one to the one reported above since for the new dataset this parameter is well constrained and its value is around $\Omega_{k0} \sim 0.1$.}. The values of the Hubble constant at present times are given in km/s/Mpc units.

\setlength{\arrayrulewidth}{0.1mm}
\setlength{\tabcolsep}{10pt}
\renewcommand{\arraystretch}{1.5}

\begin{table}[htb]
\centering
\begin{tabular}{||c | c | c | c | c ||} 
 \hline
 Dataset & $H_0$ & $\Omega_{m0}$ &  $\Omega_{k0}$ & $\sigma$ \\  
 \hline\hline
 CC \& BAO & $69.804\,^{1.878}_{-2.351}$ & $0.302\,^{0.019}_{-0.021}$ & $-0.002\,^{0.068}_{-0.066}$ & $1.121\,^{0.125}_{-0.116}$ \\
  Pantheon+ \& SH0ES  & $69.832\,^{0.334}_{-0.339}$ & $0.302\,^{0.019}_{-0.021}$ & $0.092\,^{0.025}_{-0.032}$ & $1.002\,^{0.097}_{-0.100}$  \\
   CC \& BAO \& Pantheon+ \& SH0ES & $70.266\,^{0.304}_{-0.303}$ & $0.303\,^{0.018}_{-0.021}$ & $0.092\,^{0.017}_{-0.019}$ & $1.193\,^{0.062}_{-0.077}$ \\
 \hline
\end{tabular}
\caption{Best fit values for the different datasets.}
\label{table:Best_Fit}
\end{table}

\begin{figure}[htb]
    \centering
    \hspace{1.cm}
    % First row with two minipages
    \begin{minipage}[b]{0.4\textwidth}
        \centering
    \includegraphics[width=1.\linewidth]{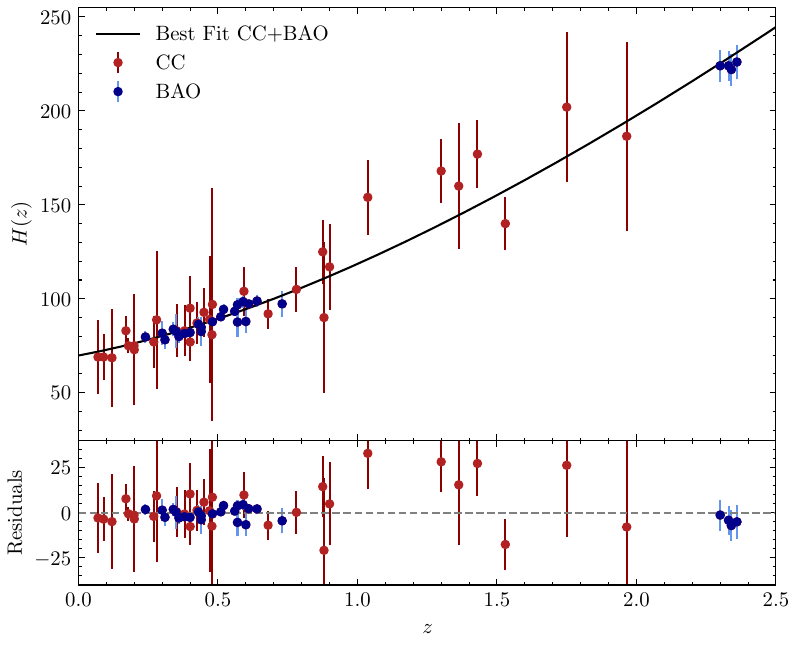}
    \end{minipage}%
    \hfill
    \begin{minipage}[b]{0.4\textwidth}
        \centering
    \includegraphics[width=1.\linewidth]{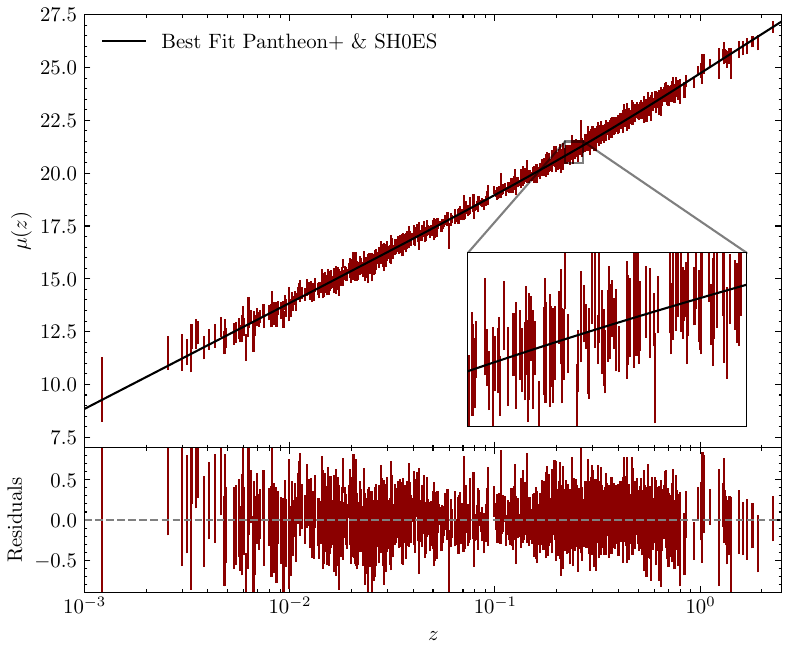}
    \end{minipage}%
    \hspace{1.cm}
    \par
    \hspace{1.cm}
    % First row with two minipages
    \begin{minipage}[t]{0.4\textwidth}
        \centering
    \caption*{Fit of $H(z)$ for the CC \& BAO datasets.}
  \label{fig:Fit_H}
    \end{minipage}%
    \hfill
    \begin{minipage}[t]{0.4\textwidth}
        \centering
   \caption*{Fit of $\mu(z)$ for the Pantheon+ \& SH0ES datasets.}
  \label{fig:Fit_mu}
    \end{minipage}%
    \hspace{1.cm}
    \caption{Best fit curves for the two different dataset for the respective fitted values of the parameters of Table \ref{table:Best_Fit}. The above plots represent the respective residuals between the theoretical model and the data.}
\label{fig:Best_fit_figures}
\end{figure}

The best fit values for the cosmological parameters obtained from the different datasets are reported in Table \ref{table:Best_Fit}. We can see from the values of $H_0$ that this model can alleviate the Hubble tension. The value of $H_0$ for the CC data is $H_\mathrm{0} = 69.804\,^{1.878}_{-2.351}$ is shifted to bigger values with respect to the standard $\Lambda$CDM case (which is $H_\mathrm{0} \sim 68 $), this in turn helps in matching this result with the SNIa data that provide a much bigger value of such constant. For the  Pantheon+ \& SH0ES compilation we see that the mean value of the Hubble parameter at present time is $H_\mathrm{0} = 69.832\,^{0.334}_{-0.339}$, which, at the contrary of the precedent case, is a smaller prediction of $H_0$ and so is more compatible with the result obtained with the other dataset. The joint analysis of CC \& BAO \& Pantheon+ \& SH0ES gives  $70.266\,^{0.304}_{-0.303}$, which is larger than that of the prediction coming from the $\Lambda$CDM case --- this provides a possible resolution of the Hubble tension issue. Moreover, the joint analysis also favors a slight positive spatial curvature of the universe. We see how these values indicate a reliable result for all the three cases, as can be also seen from Figure \ref{fig:Best_fit_figures}, where are plotted the two curves for the best fit parameters along with the respective datasets and their residuals.

\begin{figure}[htb]
    \centering
    \includegraphics[width=0.8\linewidth]{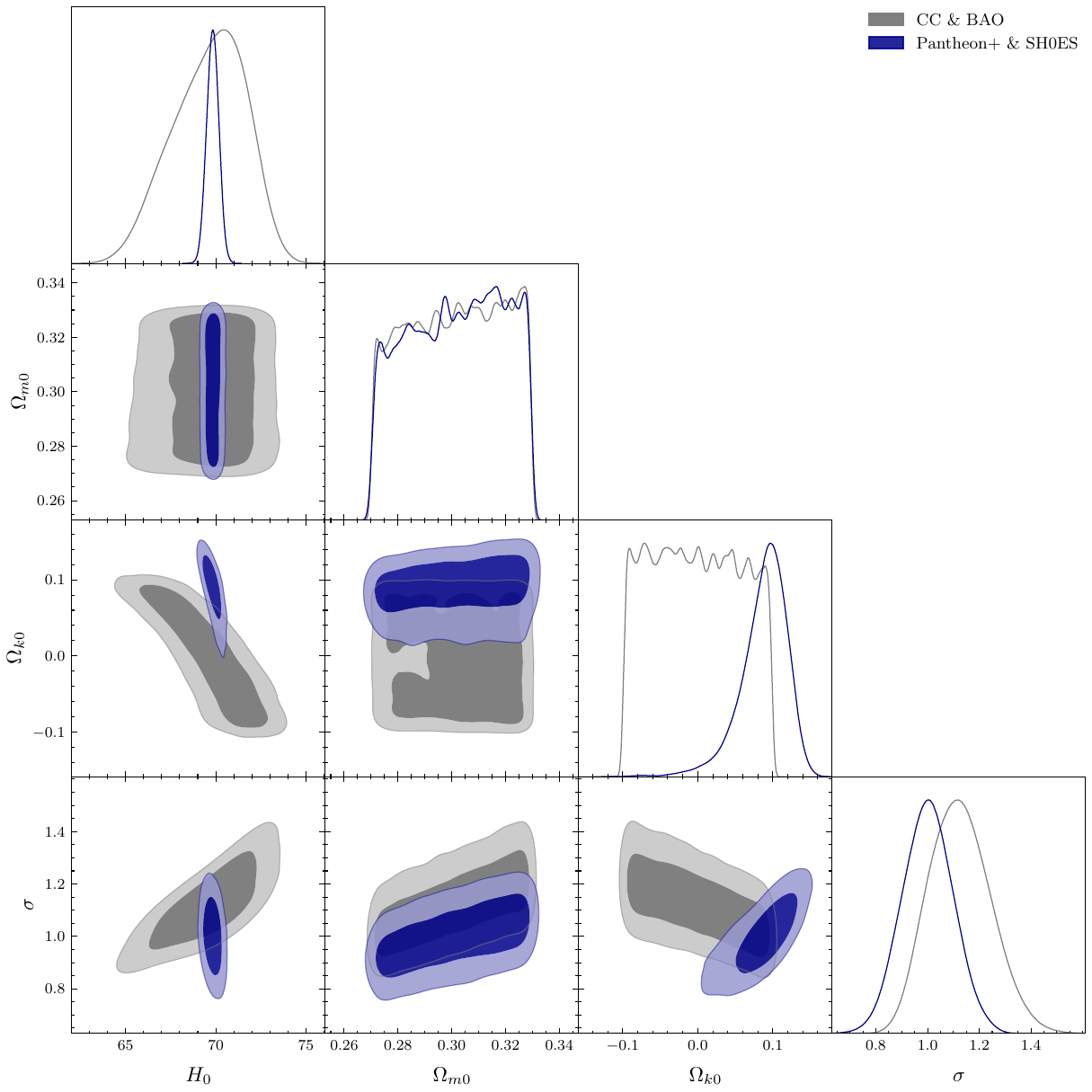}
    \caption{Corner plot for the CC \& BAO and  Pantheon+ \& SH0ES datasets analyzed separately.}
    \label{fig:CC_and_Pan-result}
\end{figure}

\begin{figure}[htb]
    \centering
\includegraphics[width=0.8\linewidth]{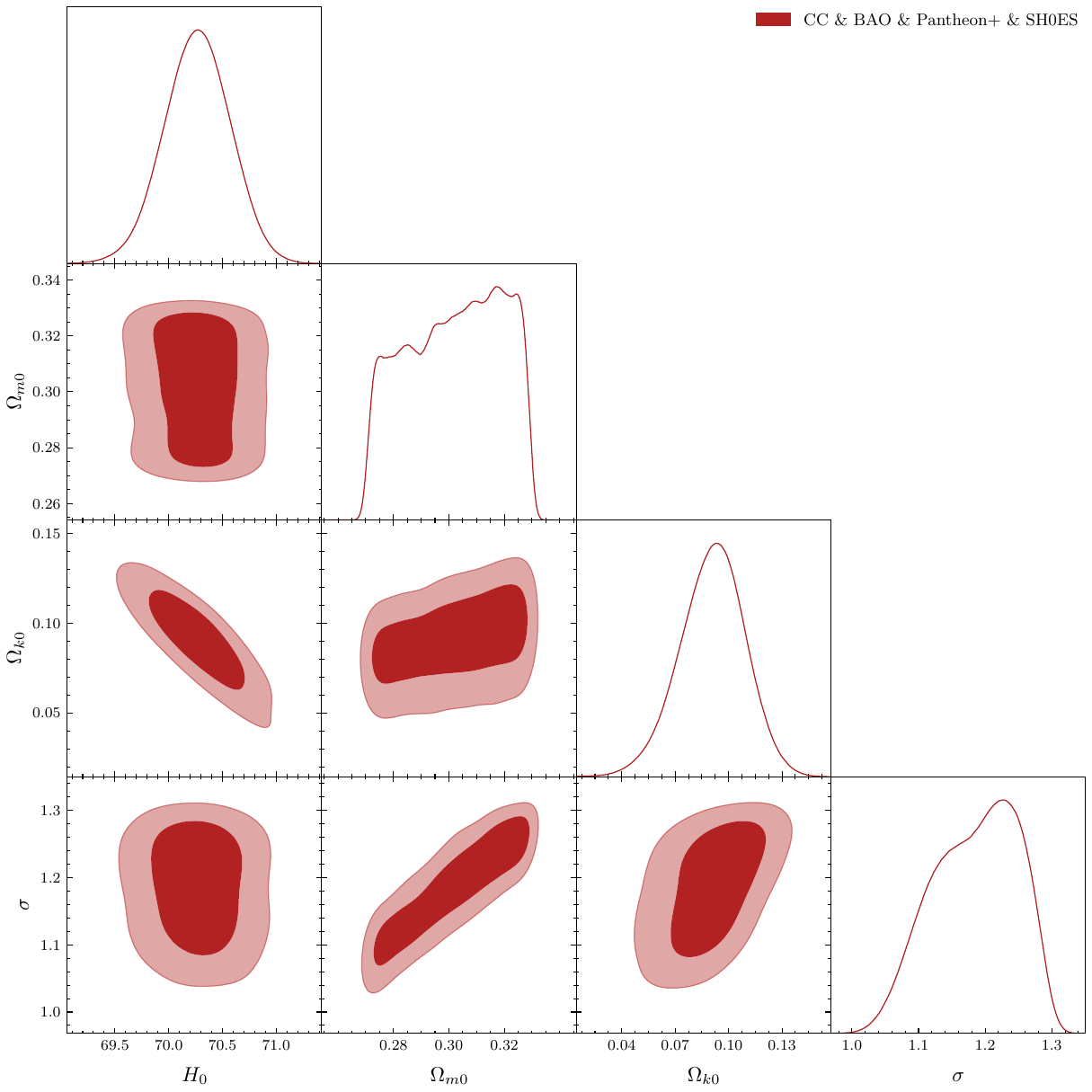}
    \caption{Corner plot for the joint analysis of the CC \& BAO \&  Pantheon+ \& SH0ES datasets.}
    \label{fig:CC+Pan-result}
\end{figure}

We also report here the different corner plot given by the MCMC analysis. Figure \ref{fig:CC_and_Pan-result} demonstrates how the model fits the two different data separately. We may note that the fit on the Pantheon+ \& SH0ES datasets is more narrow and gaussian, probably due to the larger number of data points. Also, the marginalized distribution of the entropic parameter $\sigma$ is highly gaussian, depicting a good constraint from this analysis. Regarding the spatial curvature, the $\Omega_\mathrm{k0}$ parameter is not constrained from the CC+BAO, but the SNIa data can give a good fit for this variable. In particular, the best fitted value of $\Omega_\mathrm{k0}$ comes as positive, which depicts a slight positive spatial curvature of the present universe. The analysis of the joint dataset is given in  Figure \ref{fig:CC+Pan-result}. From the plot, we still see a good tendency of the model to fit $H_\mathrm{0}$ and $\sigma$. In this case, the fit is more precise and it is also able to give reasonable result for $\Omega_\mathrm{k0}$, confirming the mean value given by the Pantheon+ \& SH0ES analysis alone. In the $\sigma - \Omega_{k0}$ plane, there seems to be a linear dependence of these two parameters, which is already present in Figure \ref{fig:CC_and_Pan-result}, but less marked. Moreover, the matter fractional density parameter, i.e. $\Omega_\mathrm{m0}$, is constrained within $\Omega_\mathrm{m0} \approx [0.28,0.32]$ from the joint analysis of CC \& BAO \& Pantheon+ \& SH0ES.

\begin{figure*}[ht]
\begin{center}
\includegraphics[width=0.49\columnwidth]{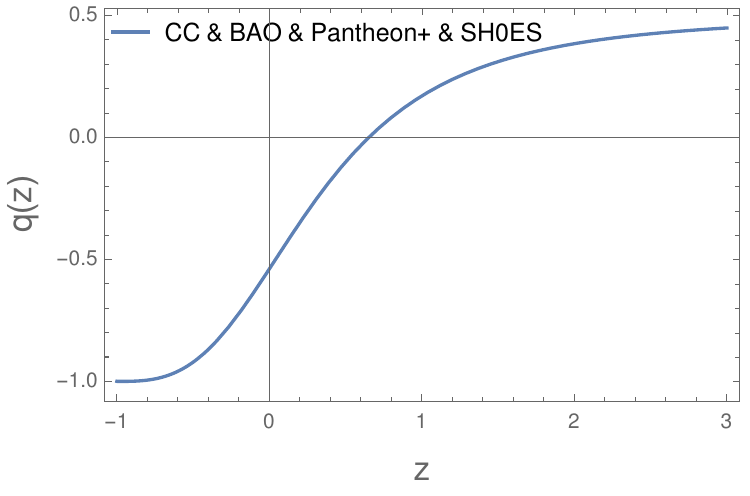}
\includegraphics[width=0.49\columnwidth]{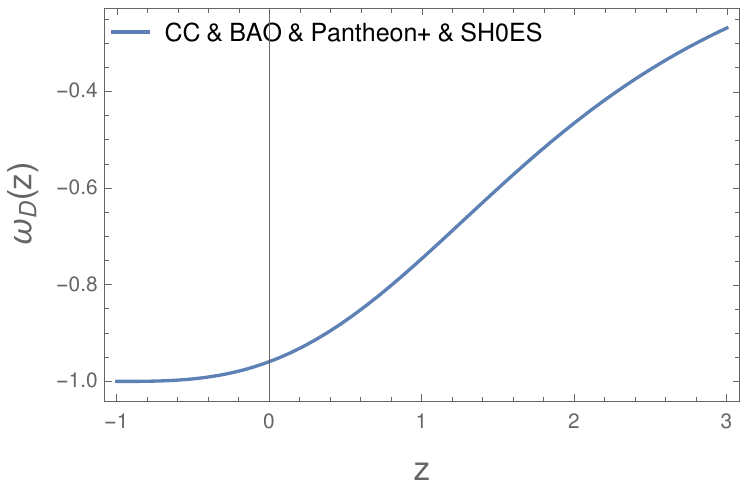}
\caption{Left Plot: Deceleration parameter, $q(z)$ vs. $z$ with the best-fitted parameters for different datasets. Right Plot: Dark energy EoS parameter, $\omega_\mathrm{D}(z)$ vs. $z$ with the best fitted parameters corresponding to the joint analysis of the CC \& BAO \&  Pantheon+ \& SH0ES datasets.}
\label{qvsz}
\end{center}
\end{figure*}

The left and right part of Fig.~[\ref{qvsz}] represent the deceleration parameter and the dark energy EoS parameter (with respect to $z$), respectively, for the best fitted values corresponding to the joint analysis of CC \& BAO \&  Pantheon+ \& SH0ES datasets (see Table.~[\ref{table:Best_Fit}]). The figure clearly demonstrates that the $q(z)$ exhibits a smooth transition from a decelerating to an accelerating universe at recent past, near about $z \sim 0.6$. Moreover, $q(z)$ approaches to $q(z) \rightarrow -1$ at far future depicting a future de-Sitter universe. Regarding the dark energy EoS parameter, it shows a non-phantom behavior during the dark energy era, and similar to the behavior of $q(z)$, $w_\mathrm{D}$ also shows a de-Sitter universe at $z \rightarrow -1$. Actually the matter density and the curvature density parameter scales as $\Omega_\mathrm{m} \sim (z+1)^3$ and $\Omega_\mathrm{k} \sim (z+1)^2$ at far future and thus they both tend to zero at $z \rightarrow -1$; this in turn makes the universe to be of a de-Sitter character during the same.
%------------------------------------
%-------------------------------------
 %-------------------------------------

\section{Conclusion}\label{Sec-conclusion}
The present work proposed a dark energy model through thermodynamic route of apparent horizon in spatially non-flat universe, where the entropy of the horizon is of 4-parameter generalized entropy. The motivation behind considering such an entropy is that it can generalize all the known entropies proposed so far in the literature for suitable representations of entropic parameters. The generalized entropy of apparent horizon induces an effective energy density in the modified Friedmann equations, which turns out to be favorable for the late time acceleration of a spatially non-flat universe. Consequently the dark energy fractional density and the dark energy EoS parameter have been found in closed analytic forms, which indeed depends on the generalized entropic parameters and the observationally determined quantities $\left\{H_\mathrm{0},\Omega_\mathrm{m0},\Omega_\mathrm{k0}\right\}$. With such dependencies, we have tested the model with latest datasets like CC \& BAO, Pantheon+ \& SH0ES and joint analysis of the CC \& BAO \&  Pantheon+ \& SH0ES respectively. For this we have carried out the $\chi^2$- minimization method and have performed the Markov Chain Monte Carlo (MCMC) analysis \citep{padilla2021cosmological} using emcee package \citep{foreman2019emcee}. It shows that the present entropic dark energy model can efficiently describe the late time cosmic acceleration preceded by a decelerated expansion phase for some best fitted values of the entropic parameters. This in turn leads to a positive spatial curvature of the present universe. Moreover the dark energy EoS parameter exhibits a non-phantom behavior at present epoch, while it converges to the $\Lambda$CDM case at far future i.e. $w_\mathrm{D} \rightarrow -1$ ar $z \rightarrow -1$. Importantly, the MCMC analysis yields the present value of the Hubble parameter as $H_\mathrm{0} \approx 70.266$ which is larger than that of in the $\Lambda$CDM scenario, and thus the current model may serve a possible resolution of the Hubble tension issue.

\appendix
\section{Computation of $\Omega_D(z)$ and $\frac{d\Omega_D}{dz}$}\label{app:Omega_D}
For the computation of $\Omega_D(z)$ and $\frac{d\Omega_D}{dz}$ we will need to rewrite the expressions of $x$ and $H^2$. Using \eqref{Def_x} and \eqref{H} we have
\begin{equation}\label{x_Tilde}
    x = \frac{H_0^2(1+z)^2}{1-\Omega_D(z)} \Tilde{\Omega}(z)\left(1+\frac{\Omega_{k0}}{\Tilde{\Omega}(z)}\left(1-\Omega_D(z)\right)\right)
\end{equation}
and 
\begin{equation}\label{H^2_Tilde}
    H^2 = \frac{H_0^2(1+z)^2}{1-\Omega_D(z)}\Tilde{\Omega}(z) \ ,
\end{equation}
where we defined for convenience the function of the red-shift
\begin{equation}
    \Tilde{\Omega}(z) \equiv (z+1)\Omega_{m0}-\Omega_{k0} \ .
\end{equation}
To find $\Omega_D$ we use the definition 
\begin{equation}
    \Omega_D \equiv \frac{8\pi G}{3 H^3} \rho_D = \frac{\Lambda}{3H^2} + \frac{x}{H^2}\left(1-\frac{\sigma}{2-\beta}x^{1-\beta}\right) \ , 
\end{equation}
then using the expression of $x$ and $H^2$, \eqref{x_Tilde} and \eqref{H^2_Tilde}, we get 
\begin{equation}
    \Omega_D = \frac{\Lambda (1-\Omega_D)}{3 H_0^2(1+z)^2\Tilde{\Omega}} + \left(1+\frac{\Omega_{k0}}{\Tilde{\Omega}}\left(1-\Omega_D\right)\right)\left(1-\frac{\sigma}{2-\beta}x^{1-\beta}\right) \ .
\end{equation}
Then we explicit $1-\Omega_D$ in this equivalence 
\begin{equation}
    \left(1-\Omega_D \right)\left(1+ \frac{\Omega_{k0}}{\Tilde{\Omega}} + \frac{\Lambda }{3 H_0^2(1+z)^2\Tilde{\Omega}}\right) = \left(1+\frac{\Omega_{k0}}{\Tilde{\Omega}}\left(1-\Omega_D\right)\right)\frac{\sigma}{2-\beta}x^{1-\beta} \ .
\end{equation}
Substituting the variable $x$ using \eqref{Def_x} and rearranging the terms we have
\begin{equation}
    \left(\frac{1+\frac{\Omega_{k0}}{\Tilde{\Omega}}(1-\Omega_D)}{1-\Omega_D}\right)^{2-\beta} = \frac{1+\frac{\Omega_{k0}}{\Tilde{\Omega}}+\frac{\Lambda}{3H_0^2(1+z)^2\Tilde{\Omega}}}{\frac{\sigma}{2-\beta}\left[H_0^2(1+z)^2\Tilde{\Omega}\right]^{1-\beta}} \ .
\end{equation}
Now we can solve this equation for $\Omega_D$ finally obtaining
\begin{equation}\label{Omega_D(z)}
    \Omega_D(z) = 1+ \left\{\frac{\Omega_{k0}}{\Tilde{\Omega}(z)}-\left[\frac{1+\frac{\Omega_{k0}}{\Tilde{\Omega}(z)}+\frac{\Lambda}{3H_0^2(1+z)^2\Tilde{\Omega}(z)}}{\frac{\sigma}{2-\beta}\left[H_0^2(1+z)^2\Tilde{\Omega}(z)\right]^{1-\beta}}\right]^{\frac{1}{2-\beta}}\right\}^{-1} \ ,
\end{equation}
where $\Tilde{\Omega}(z) \equiv (z+1)\Omega_{m0}-\Omega_{k0}$. 
To express $d\Omega_D/dz$ in a shorter expression we introduce the two functions $\mathscr{A}(z)$ and $\mathscr{B}(z)$, so that  
\begin{equation}
    \Omega_D(z) = 1+ \left\{\frac{\Omega_{k0}}{\Tilde{\Omega}(z)}-\left[\frac{\mathscr{A}(z)}{\mathscr{B}(z)}\right]^{\frac{1}{2-\beta}}\right\}^{-1} \ ,
\end{equation}
with 
\begin{equation}
    \mathscr{A}(z) \equiv 1+\frac{\Omega_{k0}}{\Tilde{\Omega}(z)}+\frac{\Lambda}{3H_0^2(1+z)^2\Tilde{\Omega}(z)} \hspace{1.cm} \text{and} \hspace{1.cm}\mathscr{B}(z)\equiv \frac{\sigma}{2-\beta}\left[H_0^2(1+z)^2\Tilde{\Omega}(z)\right]^{1-\beta} \ .
\end{equation}
The derivative of with respect to the red-shift is written as
\begin{equation}\label{Omega_D'(z)}
    \frac{d\Omega_D}{dz} = \left[\frac{\Omega_{k0}}{\Tilde{\Omega}(z)} -\left(\frac{\mathscr{A}(z)}{\mathscr{B}(z)}\right)^{\frac{1}{2-\beta}}\right]^{-2} \times \left[\frac{\Omega_{k0}\Omega_{m0}}{\Tilde{\Omega}^2}-\frac{1}{2-\beta}\left(\frac{\mathscr{A}(z)}{\mathscr{B}(z)}\right)^\frac{1}{2-\beta}\left(\frac{\mathscr{A'}(z)}{\mathscr{A}(z)}-\frac{\mathscr{B}'(z)}{\mathscr{B}(z)}\right)\right] \ ,
\end{equation}
where 
\begin{equation}
    \frac{\mathscr{A}'(z)}{\mathscr{A}(z)} = -\frac{1}{\Tilde{\Omega}}\frac{\Omega_{k0}\Omega_{m0}-\frac{\Lambda}{3H_0^2(z+1)^4}\left[3(1+z)^2\Omega_{m0}-2(1+z)\Omega_{k0}\right]}{\Tilde{\Omega}+\Omega_{k0}+\frac{\Lambda}{3H_0^2(1+z)^2}}
\end{equation}
and
\begin{equation}
   \frac{ \mathscr{B}'(z)}{ \mathscr{B}(z)} = \frac{1-\beta}{(1+z)^2\Tilde{\Omega}}\left[3(1+z)^2\Omega_{m0}-2(1+z)\Omega_{k0}\right] \ .
\end{equation}

\section*{Acknowledgments}
This work is funded by MCIN/AEI/10.13039/501100011033 and FSE+, reference PRE2021-098098 (S. D'Onofrio) and the program Unidad de Excelencia María de Maeztu
CEX2020-001058-M (S.D. Odintsov and S. D'Onofrio).

\phantomsection

\bibliography{GEN}

\bibliographystyle{apsrev4-2}

\end{document}